\documentstyle[osa,epsfig,times,twocolumn]{revtex}

\begin{document}

\title{How big is a quantum computer?}

\author{S. Wallentowitz,$^{1,2}$ I. A. Walmsley$^{1,3}$ and J. H.
Eberly$^{1,2,3}$}

\address{$^1$Center for Quantum Information, University of Rochester,
  Rochester, New York 14627, USA \\
  $^2$Rochester Theory Center for Optical Science and Engineering and
  Department of Physics and Astronomy, University of
  Rochester, Rochester, New York 14627, USA \\
  $^3$The Institute of Optics, University of Rochester, Rochester, New
  York 14627, USA}

\maketitle


{\bf Accounting for resources is the central issue in computational
  efficiency.  We point out physical constraints implicit in
  information readout that have been overlooked in classical
  computing. The basic particle-counting mode of read-out sets a lower
  bound on the resources needed to implement a quantum computer. As a
  consequence, computers based on classical waves are as efficient as
  those based on single quantum particles.}

Every quantum information processor must be coupled to a decoder, a
device that condenses the contents of the quantum register into the
classical information that makes up the computational output. The
details of the decoder itself can have a profound effect on the power
of a given quantum information processor. In the laboratory, decoding
is usually done by particle counting, using a detector with finite
spatial, temporal, spectral, etc., resolution -- and this has some
significant implications for the resources that may be needed to
implement a quantum processor.

The formal theory of processor operation (the quantum state-space
algebra) does not sharply distinguish between a quantum information
processor consisting of a few particles occupying many states or many
particles occupying a few states. But obviously the decoder (particle
counter) can make the distinction easily. The distinction is closely
related to the different kinds of entanglement central to advantages
accrued by working in the quantum domain. For a single particle,
entanglement is usually defined in terms of states belonging with
different degrees of freedom of the particle (and associated with
commuting observables). For many-particle systems entanglement is
possible in addition between identical degrees of freedom of different
particles. In any case a particle-counting register readout simply
provides a dichotomic answer to the question as to whether there was a
particle in a particular mode of the system as a whole.

Information processing by photons provides an illustration. A typical
apparatus contains photodetectors whose function is to count by
photoionization the number of photons in a certain spatial and
spectral range. The photon-counting detector can therefore be
understood to give a direct answer to the question: How many photons
are in detector mode $\lambda$? Note that the mode label $\lambda$
refers to a set of numbers specifying the modal character. For a
photon, these might be the frequency, polarization and wavevector. A
different and equally uncomplicated example is provided by the
measurement of the electronic state of a Rydberg atom. Standard
ramped-field ionization detection gives an answer to the question: Is
a particular electronic state, representing an eigenmode of the
Schr\"odinger equation, occupied or not?  In fact, because of the
particle-counting register readout, a quantum information processor
that is implemented by means of the modal entanglement of a single
particle can be implemented with equal efficiency by classical wave
interference.  It may be that detection schemes that do not rely on
particle counting could show improvements beyond those of classical
wave interferometers.  However, we are aware of no laboratory schemes
for implementing such measurements.

Any von-Neumann type measurement requires for each of $N$ eigenvalues
of interest a distinct measurement capability.  In the simplest
example one can envisage these might be the $N$ channels in a Zeeman
or Stern-Gerlach apparatus.  For such a case, one says that a
``resource'' of order $N$ must be on
hand\cite{zeilinger,barenco,kwiat1,lloyd,kwiat2}. With this notion of
a measurement and its resources, we can quantify the ``resource
space'' needed to implement a combination of processor and decoder. It
is simply the total spatio-temporal volume of the modes
defining the measurement capabilities. This aspect of the resource
requirement is usually ignored in classical computer science, because it is
assumed that such a volume may be infinitesimal. Yet this is
completely unphysical: quantum mechanics requires that the phase-space
volume associated with each mode must be at least $\hbar$, and consequently
the spatio-temporal volume of a mode cannot be zero for a system with finite
energy. In addition to counting the resources required for the processor,
quantum mechanics therefore forces us to reassess the resources required for
readout.

In the specific case of a single-particle processor, the quantum
register that is processed during the computation is expressed by a
quantum state in which the particle occupies a superposition of $N$
different modes. The readout of such a register then requires at least
$\log_2 \! N$ distinct detectors or generalized resources. The
classical outputs of these detectors then have to be polled to locate
the sequence of detectors that fired. This is simply the classical
search problem for a sorted database, and it is evident that such a
search can be done most efficiently in $\log_2\! N$ steps by means of
a binary search. This method of resource counting can be adopted to
show that any single-particle processor can be efficiently implemented
using classical wave interference, even when the system relies
exclusively on the entanglement of different degrees of freedom of the
particle.

\begin{figure}
  \begin{center}
    \epsfig{file=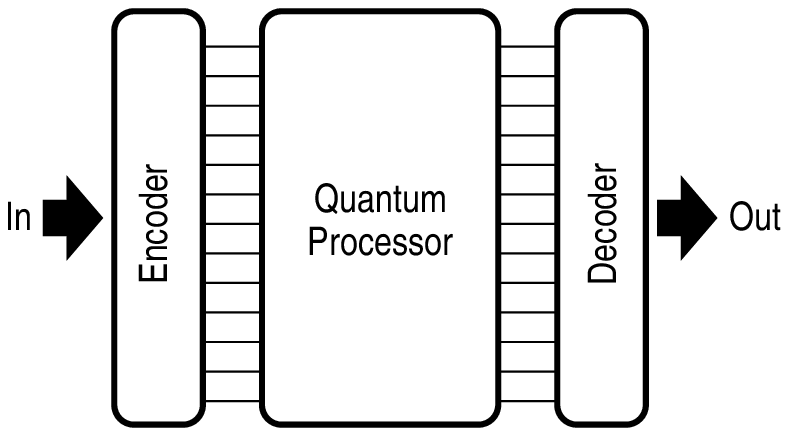,scale=1}
  \end{center}
  {\bf Figure~1} Elements of a quantum computer. The encoder
  transcribes the classical input into the quantum register of the
  processor. After completion of the processing, the quantum register
  is read out by the decoder to obtain the classical computational
  output.
  \label{fig:cpu}
\end{figure}

A description of the quantum processor in terms of the system state
vector alone makes it difficult to distinguish single-particle from
many-particle situations, let alone to understand the optimal
combinations of particles and degrees of freedom that might be needed
to accomplish a certain processing task. The essential roles of
interference and entanglement can be examined unambiguously most
simply in the language of second quantization.  The field operator
$\hat{\psi}(x)$ describes the annihilation of atoms, photons or other
types of particles at position $x$. If we introduce a complete set of
orthonormal spatial modes $u_\lambda(x)$, where $\lambda$ labels the
different modes including the possibility of having several degrees of
freedom, we may introduce mode annihilation operators
$\hat{a}_\lambda$ via $\hat{\psi}(x) \!=\! \sum_\lambda
\hat{a}_\lambda u_\lambda(x)$, where $[\hat{a}_\lambda,
\hat{a}^\dagger_\mu]_\pm \!=\!  \delta_{\lambda\mu}$ depending on
whether the field describes fermions or bosons.

If the unitary operator $\hat{U}$ defines the evolution of a quantum
processor, then the output register is given by
\begin{equation}
  \label{eq:unitary-evol}
  \hat{\psi}(x,t) = \hat{U}(t) \, \hat{\psi}(x,0) \, \hat{U}^\dagger
  (t) ,
\end{equation}
where $\hat{\psi}(x,0)$ represents the input register. The only
restriction on the Hamiltonian generating $\hat{U}$ is that it
preserves the particle number of the processor register: If this
register contains only a single particle, so that the state is given
by $|\psi\rangle \!=\! \sum_\lambda \psi_\lambda
\hat{a}_\lambda^\dagger |{\rm vac}\rangle$, the only terms in the
normally ordered Hamiltonian contributing to the dynamical evolution
are those bilinear in $\hat{a}_\lambda^\dagger$ and $\hat{a}_\lambda$.
Thus the unitary evolution (\ref{eq:unitary-evol}) reduces from the
operators to the mode functions themselves:
\begin{eqnarray}
  \label{eq:proc-dynamics}
  \hat{\psi}(x,t) = \sum_\lambda \hat{a}_\lambda u_\lambda(x,t) ,
\end{eqnarray}
where $u_\lambda(x,t)$ are the time-propagated mode functions of wave
mechanics.

The operation of an elementary measurement can be formulated in the
language of fields as the action of a linear filter on the underlying
mode structure of the system. The register field operator after the
filter is given by
\begin{equation}
  \label{eq:mode-filtering}
  \hat{\psi}_\mu(x,t) = \int \! dx' \! \int \! dt' \, \Gamma_\mu(x,t|x',t')
  \, \hat{\psi}(x',t') ,
\end{equation}
where $\mu$ denotes the mode that is to be probed for particles.  The
observable corresponding to an elementary measurement is given by the
number operator for this mode
\begin{equation}
  \label{eq:intensity}
  \hat{N}_\mu = \int \! dx \! \int \! dt \, \hat{\psi}_\mu^\dagger(x, t)
  \, \hat{\psi}_\mu(x,t) .
\end{equation}
In each measurement one determines the number of particles in the
specified detector mode. This can be described by the projector
$|N;\mu \rangle\langle N;\mu| \!=\! \hat{\delta}(\hat{N}_\mu \!-\!
N)$, where $|N;\mu\rangle$ denotes the state with $N$ particles in the
detected mode $\mu$ and $\delta(N)$ is understood as the Kronecker
delta function.  In the case of a single particle, $N$ may take the
values $(0,1)$, so that the projector simplifies to
\begin{equation}
  \label{eq:sp-projector}
  \hat{\delta}(\hat{N}_\mu \!-\! N) =  \delta(N) \, ( \hat{1}_\mu
  \!-\! \hat{N}_\mu ) + \delta (N \!-\! 1) \, \hat{N}_\mu ,
\end{equation}
showing that the measurement projector can be entirely expressed in
terms of the bilinear number operator~(\ref{eq:intensity}).

As indicated previously, the minimum number of detectors required to
read out the $N$-channel register is $\log_2 \! N$. This can be
achieved for any quantum system by representing each of the mode
labels $\mu$ by a set of $\log_2 \! N$ bits $\{ b_i \}$ and grouping
the number operators of the modes, $\hat{N}_{\{b_i\}}$, into a
specific set of detectable number operators $\hat{M}_\alpha$ according
to their common bits. For example, in a system of $N\!=\!8$ different
modes (i.e. with a Hilbert space of eight dimensions) the three number
operators are
\begin{eqnarray}
  \label{eq:number-op-binary}
  \hat{M}_{1xx} & = & \hat{N}_{\{1,0,0\}} + \hat{N}_{\{1,0,1\}} +
  \hat{N}_{\{1,1,0\}} + \hat{N}_{\{1,1,1\}} , \\
  \hat{M}_{x1x} & = & \hat{N}_{\{0,1,0\}} + \hat{N}_{\{0,1,1\}} +
  \hat{N}_{\{1,1,0\}} + \hat{N}_{\{1,1,1\}} , \\
  \hat{M}_{xx1} & = & \hat{N}_{\{0,0,1\}} + \hat{N}_{\{0,1,1\}} +
  \hat{N}_{\{1,0,1\}} + \hat{N}_{\{1,1,1\}} .
\end{eqnarray}
The readout would then occur via a cascaded sequence of particle
counters. The result of measuring the three
observables~(\ref{eq:number-op-binary}) is the set of bits specifying
the mode. Here too the projectors corresponding to the operators
$\hat{M}_\alpha$ take the form~(\ref{eq:sp-projector}). Note that it
is \underline{not} necessary that each of the bits correspond to a
different degree of freedom of the particle, i.e. $\mu \!=\!
\{\mu_1,\mu_2,\ldots\}$. The binary readout scheme is possible even
for a single particle excited in one degree of freedom. In practice it
may prove simpler to use different degrees of freedom (such as the
principle quantum number, and the two angular momentum quantum numbers
for a Rydberg electron), though no such scheme has been analyzed in
detail as yet.  Though the number of detectors is minimal, the number
of resources in terms of space-time volume of required modes is not,
as can be seen from the fact that each measured number operator
involves more than one mode.

If the result of the computation leaves the register in one of the
detected modes, then the readout provides directly a classical result
for the computation. In general, however, the computation will place
the register in a superposition state, so that the readout will be
probabilistic. That is, for many runs of the same computation the
classical outputs will be different. Therefore the whole computational
process has to be repeated many times, either sequentially or in
parallel. Thus in both cases one must determine for each mode the
expectation value of the projector (\ref{eq:sp-projector}).

For a single particle it is straightforward to show using
Eqs~(\ref{eq:mode-filtering}) -- (\ref{eq:sp-projector}) that this
expectation value is proportional to the correlation function
\begin{equation}
  \label{eq:avg-population}
  C(x,t;x',t') = \langle \hat{\psi}^\dagger (x,t) \, \hat{\psi}(x',t')
  \rangle .
\end{equation}
Using the general single-particle state $|\psi \rangle \!=\!
\sum_\lambda \psi_\lambda \hat{a}_\lambda^\dagger | {\rm vac} \rangle$
and the dynamics according to Eq.~(\ref{eq:proc-dynamics}) we obtain
\begin{equation}
  \label{eq:exp-value}
  C(x,t;x',t') = \psi^\ast(x,t) \, \psi(x',t') ,
\end{equation}
where we have defined the single-particle wavefunction as $\psi(x,t)
\!=\!  \sum_\lambda \psi_\lambda u_\lambda(x,t)$. This result holds
for both the single-atom and single-photon case and is equivalent to a
classical interference pattern obtained by replacing the expectation
value~(\ref{eq:exp-value}) by the classical electric-field correlation
function
\begin{equation}
  \label{eq:cl-interference}
  C_{\rm class}(x,t;x',t') = E^\ast(x,t) \, E(x',t') .
\end{equation}
where $E(x,t)$ is an analytic signal. A comparison of
Eqs~(\ref{eq:exp-value}) and (\ref{eq:cl-interference}) shows a
complete one-to-one correspondence between classical interference and
single-particle quantum interference, not only in the dynamics\cite{jozsa}
but also in the measurement. We conclude therefore that any quantum
computer based on a single-particle quantum register can be
implemented {\it with equal efficiency} entirely by a classical
interferometer. This is because the concept of the entanglement of
degrees of freedom of a single particle cannot be attributed an
inherently quantum character: it is perfectly understandable in terms
of classical wave interference.

This formulation also provides some insight into the issue of the
speedup of a particular algorithm using quantum interference.
Consider, for example, a quantum search algorithm that is based on
single-particle entanglement, that is, on the correlation of different
degrees of freedom of a single particle at the level of probability
amplitudes. In this case the quantum register may be realized by $N
\gg 1$ modes but the modes can only be occupied by a total population
of 1.  In the optical case, these might be the modes associated with
the polarization and the wavevector degrees of freedom\cite{kwiat2}.
Alternatively the energy and angular momentum degrees of freedom of a
Rydberg electron may be used\cite{bucksbaum}.

The readout requires $O(N)$ resources and at minimum $\log_2 \! N$
detectors, whose classical outputs can be searched in not less than
$O(\log_2 \!  N)$ steps. In fact, it is evident that a classical
search is always required in the final readout, since the values of
the output quantum register only obtain physical reality after having
been measured.  Therefore, any quantum computation with a single
particle cannot be faster than $O(\log_2 \! N)$. Moreover, by virtue
of the one-to-one correspondence described above, it can be performed
using classical interference with the same speed and the same number
of resources.  This is in striking contrast to the belief that a
quantum-search algorithm, even without using multi-particle
entanglement, i.e. with only a single particle, can be advantageous
for solving the search problem\cite{knight}.

The quantum computation itself consists of preparing the input
register and performing a unitary $N \!\times\! N$ transformation
$\hat{U}$. According to Grover\cite{grover1}, this unitary
transformation may perhaps consist of a series of operations like
passing the single particle through a so-called Oracle, whose function
is to alter the phase of one of the modes, followed by an $N
\!\times\! N$-port beamsplitter. These steps are usually referred to
as ``querying the Oracle'' and ``inversion about the mean''.  Using
these operations it has been shown that an unsorted database can be
searched by performing only $O(\sqrt{N})$ ``queries'' or even only a
single ``query'' of the Oracle\cite{grover1,grover2}. This is in stark
contrast to the corresponding classical search which requires $O(N)$
queries.

The caveat, as Steane has pointed out\cite{steane}, is that one should
be careful of comparing an inefficient classical algorithm with an
efficient quantum one. The experimental implementations of the Grover
search to date\cite{kwiat2,bucksbaum} have not in fact implemented a
search of an unsorted database, but rather a database with one single
marked item.  Therefore experimental evidence for the claimed speedup
has yet not been achieved, since a sorted database can be searched
classically in $\log_2\!  N$ steps, which is identical to the readout
limit for a quantum computer.

The superiority of the quantum search algorithm becomes apparent only
when one examines carefully the notion of a query. It is evident that
there is no information gain in the sort of oracle query described
above, which involves only unitary transformations. It is in the
encoding of such oracles, which must be done by a quantum
computer\cite{boyer}, that the real speedup occurs. Once a properly
encoded oracle is available, the number of steps required to perform
the information processing is then limited by the final register
readout to $O(\log_2 \! N)$.  This is exactly the same as for a
classical information processor, which shows that it is the
realization of actual information by the readout, as opposed to
predictive information that is contained in the quantum
state\cite{heisenberg}, that is the ultimate limiting procedure in
quantum information processing.

We have shown here that if the readout of the register is performed
by particle counting then there exists a one-to-one
correspondence of single-particle quantum interference and classical
interference. Therefore we conclude that any enhancements in processing
power that can be ascribed to quantum interference can also be found in
classical wave processors, and this includes systems based on modal
entanglement. Multi-particle
entanglement, on the other hand, may provide enhancements that cannot be
efficiently transcribed to classical interferometers, even when
particle counting is used to realize the output information.

\noindent {\bf Acknowledgments} \\[0.5ex]
This work was supported by an ARO-administered MURI Grant No.
DAAG-19-99-1-0125 (IAW and JHE), and by NSF Grant No. PHY-9415583
(JHE). SW acknowledges support by the Studienstiftung des deutschen
Volkes. We acknowledge several stimulating interactions with P.
Bucksbaum, P.L. Knight, M.G. Raymer, C.R. Stroud, Jr. and K.
Wodkiewicz.

\vspace*{2ex}

\noindent Correspondence should be addressed to I. A. Walmsley (email:
walmsley@optics.rochester.edu).



\begin{references}

\bibitem{zeilinger} Reck, M., Zeilinger, A., Bernstein, H.J. \&
  Bertani, P.  Experimental realization of any discrete unitary
  operator. {\it Phys.~Rev.~Lett.\/} {\bf 73}, 58 (1994).

\bibitem{barenco} Barenco, A. {\it et al.\/} Elementary gates for
  quantum computation. {\it Phys.~Rev.~A} {\bf 52}, 3457 (1995).

\bibitem{kwiat1} Cerf, N.J., Adami, C. \& Kwiat, P.G. Optical
  simulation of quantum logic. {\it Phys.~Rev.~A} {\bf 57}, R1477
  (1998).

\bibitem{lloyd} Lloyd, S. Quantum search without entanglement. {\it
    Phys.~Rev.~A} {\bf 61}, 010301(R) (1999).
  
\bibitem{kwiat2} Kwiat, P.G., Mitchell, J.R., Schwindt, P.D.D. \&
  White, A.G. Grover's search algorithm: an optical approach. {\it
    J.~Mod.~Opt.\/} {\bf 47}, 257 (2000).

\bibitem{jozsa} Josza, R. Entanglement and quantum computing. {\it
    Geometric issues in the foundation of science} eds. Huggett, S.,
  Mason, L., Tod, K.P., Tsou, S.T. \& Woodhouse, N.M.J. (Oxford
  University Press, 1997).

\bibitem{bucksbaum} Ahn, J., Weinacht, T.C. \& Bucksbaum, P.H.
  Information storage and retrieval through quantum phase. {\it
    Science} {\bf 287}, 463 (2000).

\bibitem{knight} Knight, P.L. Quantum information processing without
  entanglement. {\it Science} {\bf 287}, 441 (2000).

\bibitem{grover1} Grover, L.K. Quantum mechanics helps in searching
  for a needle in a haystack. {\it Phys.~Rev.~Lett.\/} {\bf 79}, 325
  (1997).

\bibitem{grover2} Grover, L.K. Quantum mechanics can search
  arbitrarily large databases by a single query. {\it
    Phys.~Rev.~Lett.\/} {\bf 79}, 4709 (1997).

\bibitem{steane} Steane, A. A quantum computer only needs one
  universe. Los Alamos National Laboratory e-print quant-ph/0003084
  (2000).

\bibitem{boyer} Boyer, M., Brassard, G., Hoyer, P. \& Tapp, A. Tight
  bounds on quantum searching. {\it Fortsch.~Phys.\/} {\bf 46}, 493
  (1998).

\bibitem{heisenberg} Heisenberg, W. {\it The Physical Principles of
    the Quantum Theory} (University of Chicago Press, Chicago, 1930).

\end{references}
\end{document}